\documentclass[prd,amssymb,amsmath,aps,floatfix,nofootinbib, preprintnumbers ,12pt]{revtex4-1}
\pdfoutput=1

\usepackage{color}
\usepackage{graphicx}
\usepackage{setspace} 

\def\beq{\begin{equation}}
\def\eeq{\end{equation}}

\def\be{\begin{equation}}
\def\ee{\end{equation}}
\def\bea{\begin{eqnarray}}
\def\eea{\end{eqnarray}}

\newcommand{\gsim}{\lower.7ex\hbox{$\;\stackrel{\textstyle>}{\sim}\;$}}
\newcommand{\lsim}{\lower.7ex\hbox{$\;\stackrel{\textstyle<}{\sim}\;$}}

\begin{document}

\preprint{IPMU 10-0036}
\preprint{NSF-KITP-10-026}

\bigskip

\title{Entropic Accelerating Universe}

\author{Damien A. Easson$^{1,2,3}$}
\email{easson@asu.edu}
\author{Paul H. Frampton$^{1,4}$}
\email{frampton@physics.unc.edu}
\author{George F. Smoot$^{1,5,6,7,8}$}
\email{gfsmoot@lbl.gov}
\affiliation{$^1$Institute for the Physics and Mathematics of the Universe,
University of Tokyo, Kashiwa, Chiba 277-8568, Japan}
\affiliation{$^2$ Department of Physics \& School of Earth and
Space Exploration \& Beyond Center,
Arizona State University, Phoenix, AZ 85287-1504, USA}
\affiliation{ $^3$Kavli Institute for Theoretical Physics, University of California, Santa Barbara, CA 93106-4030, USA}
\affiliation{ $^4$Department of Physics and Astronomy,
University of North Carolina, Chapel Hill, NC 27599, USA}
\affiliation{$^5$Lawrence Berkeley National Lab, 1 Cyclotron Road, Berkeley, CA 94720, USA}
\affiliation{$^6$Physics Department, University of California, Berkeley, CA 94720, USA}
\affiliation{$^7$Institute for the Early Universe, Ewha Womans University \& Advanced Academy, Seoul, Korea}
\affiliation{$^8$Chaire Blaise Pascale, Universite Paris Denis Diderot, Paris}

\begin{abstract}
To accommodate the observed accelerated expansion of the universe, one
popular idea is to invoke a driving term in the Friedmann-Lema\^{i}tre
equation of dark energy which must then comprise
70\% of the present cosmological energy density. We propose
an alternative interpretation which takes into account 
the entropy and temperature intrinsic to the horizon of the universe due to the information holographically 
stored there. 
Dark energy is thereby obviated and the acceleration is due
to an entropic force naturally arising from the information
storage on the horizon surface screen.
We consider an additional quantitative approach inspired by surface terms  in general relativity and show that this leads to the entropic accelerating universe.
\end{abstract}


\maketitle

\section{Introduction}

\bigskip
\bigskip

\noindent The most important observational advance in cosmology since 
the early studies of cosmic expansion in the 1920's was the dramatic and unexpected
discovery, in the
waning years of the twentieth century, that the expansion rate is accelerating.
This was first announced in February 1998, based on the concordance of
two groups' data on Supernovae Type 1A \cite{Perlmutter98, Reiss98}.

\bigskip

\noindent A plethora of subsequent experiments concerning
the Cosmic Microwave Background (CMB), Large Scale Structure (LSS),
and other measurements have all
confirmed the 1998 claim for cosmic acceleration.
There have been many attempts to avoid the conclusion of the cosmic acceleration.
Typically they involve an ingenious ruse which 
assigns a special place to the Earth in the Universe, in a frankly
Ptolemaic manner and in contradiction to the well-tested
and time-honored cosmological principle at large distance.
We find these to be highly contrived and {\it ad hoc}.

\bigskip

\noindent We therefore adopt the position that the accelerated expansion
rate is an observed fact which we, as theorists, are behooved
to interpret theoretically with the most minimal set
of additional assumptions.

\bigskip

\section{Interpretation as Dark Energy}

\bigskip

\noindent On the basis of general relativity theory, together
with the cosmological principle of homogeneity and isotropy,
the scale factor $a(t)$ in the FRW metric satisfies \cite{F,L}
the Friedmann-Lema\^{i}tre equation

\begin{equation}
H(t)^2 = \left( \frac{\dot{a}}{a} \right)^2 = \left( \frac{8 \pi G}{3} \right) \rho  
\label{FLequation}
\end{equation}

\bigskip

\noindent where we shall normalize $a(t_0) = 1$ at the present, time $t=t_0$,
and $\rho$ is an energy density source which drives the
expansion of the universe. Two established contributions to $\rho$
are $\rho_m$ from matter (including dark matter) and $\rho_{\gamma}$
radiation, so that

\begin{equation}
\rho \supseteq \rho_{m} + \rho_{\gamma}
\label{rho}
\end{equation}

\noindent with  $\rho_{m}(t)  =  \rho_{m}(t_0) a(t)^{-3}$
and $\rho_{\gamma} (t)  = \rho_{\gamma}(t_0) a(t)^{-4}$.

\bigskip

\noindent For the observed accelerated expansion, the most  popular
approach is to add to the sources, in Eq.(\ref{FLequation}), a
dark energy term $\rho_{DE}(t)$ with 

\begin{equation}
\rho_{DE} (t) = \rho_{DE}(t_0) a(t)^{-3 (1 + \omega)}
\label{rhoDE}
\end{equation}

\bigskip
\bigskip
\noindent where $\omega = p/\rho c^2$ is the equation of state parameter. For the
case $\omega = -1$, as for a cosmological constant, $\Lambda$,
and discarding the matter and radiation terms which are relatively negligible
we can easily
integrate the Friedmann-Lema\^{i}tre equation to find 

\begin{equation}
a(t) = a(t_0) ~ e^{ H t}
\label{CC}
\end{equation}

\bigskip

\noindent where $\sqrt{3} H= \sqrt{\Lambda} = \sqrt {8 \pi G \rho_{DE}}$.

\bigskip

\noindent By differentiation of Eq. (\ref{CC}) with respect
to time $p$ times we obtain for the $p^{th}$
derivative

\begin{equation}
\frac{d^p}{dt^p} a(t) |_{t=0} = (H)^p
\label{jerk}
\end{equation}

\bigskip

\noindent Therefore, if $\Lambda > 0$ is positive,
as in a De Sitter geometry, not only is the
acceleration ($p=2$) positive and non-zero,
but so are the jerk ($p=3$), the snap ($p=4$),
the crackle ($p=5$), the pop ($p=6$) and
all $p \ge 7$.

\bigskip

\noindent The insertion of the dark energy term
Eq.~(\ref{rhoDE}) in Eq.~(\ref{FLequation})
works very well as a part of the $\Lambda CDM$
model. However, it is an {\it ad hoc} procedure
which gives no insight into what dark energy is.  Identifying the cosmological constant with vacuum energy leads to
the infamous cosmological constant problem: the observed value of the cosmological constant
$\rho_{\Lambda_{\rm obs}} \sim (10^{-3} \, \rm eV)^4$ and the theoretical prediction (assuming a UV cutoff  at the Planck scale)
$\rho_{\Lambda_{\rm th} }\sim (10^{18} \, \rm GeV)^4$, disagree by an embarrassing 120 orders of magnitude. 

\bigskip

\noindent  With this background, we shall now move
to a different explanation for the accelerated
expansion which obviates the need for any ambiguous dark energy component, including
scalar fields. This approach (which does not directly solve the cosmological constant problem) even leads to new insight into the perplexing ratio 
$\rho_{\Lambda_{\rm obs}}/\rho_{\Lambda_{\rm th}} \sim 10^{-120}$ discussed above. 

\section{Interpretation as Entropic Force}\label{sec:ent}

\bigskip

\noindent We now adopt a different approach,
with no dark energy, where instead the
central role is played by the ideas of 
information and holography,
entropy and temperature \footnote{{The entropy of the universe has received some recent attention~\cite{Lineweaver,PHF},
in  part because it relates to the feasibility of constructing a consistent cyclic model.
For example, the cyclic model in~\cite{BF}, assuming its internal consistency will indeed
be fully confirmed, may provide a solution to the difficult entropy question
originally posed  seventy-five years earlier by Tolman~\cite{Tolman}.}}.

\bigskip

\noindent The first and only assumption is that a horizon
has both a temperature and entropy associated with it. 
This was first shown clearly to hold for black hole horizons with
a temperature given by the Hawking temperature and an entropy given 
by the Bekenstein entropy.
Here we take  the apparent horizon of the universe
\footnote{In the following discussion we consider a flat $k = 0$ universe so the apparent horizon coincides with the Hubble radius. We relegate a more general discussion of $k \ne 0$ to the Appendix.}.
\bigskip

\noindent At this horizon, there is a horizon
temperature, $T_{\beta} $, which we can estimate as

\begin{equation}
T_{\beta} = \frac{\hbar}{  k_B }~\frac{H }{ 2 \pi} \sim 3 \times 10^{-30} K 
\label{T-beta}
\end{equation}

\bigskip

\noindent Such a temperature is closely related to
the de Sitter temperature~\footnote{We suspect, without
rigor, that in the third law of thermodynamics
the notion of absolute zero, $T=0$, must be replaced by 
$T \geq T_{\beta}$, although this is not our present 
concern.}.
More relevant to the central question is the fact
that the temperature of the horizon 
leads to the concomitant entropic force
and resultant acceleration $a_{Horizon}$ of the horizon given
by the Unruh~\cite{unruh} relationship

\bigskip

\begin{equation}
a_{Horizon} = \left( \frac{2 \pi c k_B T_{\beta}}{\hbar} \right) = c H \sim 10^{-9} \, m/s^2
\,
\label{acceleration}
\end{equation}

\bigskip

\noindent When $T_{\beta}$ is used in Eq. (\ref{acceleration}),
we arrive at a cosmic acceleration essentially in
agreement with the observation.

\bigskip

\bigskip

\noindent From this viewpoint, the ambiguous dark energy component is non-existent.
Instead there is an entropic force contribution acting at the horizon
and pulling outward towards the horizon to create the appearance of a dark energy component~\footnote{
The possibility that cosmic acceleration can be described by an entropic force should be distinguished from the idea
that gravity itself is an entropic force  \cite{paddy,verlinde}; although the two ideas are not \it prima facie \rm incompatible.}. 
We emphasize
again that because we have nothing new to say about the physics governing quantum fluctuations, the above
argument does not, by itself, solve the cosmological constant problem. However, the interpretation of cosmic
acceleration as due to entropic force helps to understand why the accelerating component is expected to
be small today (of order $H$ as in Eq.~(\ref{acceleration})), in contrast to the embarasingly large value predicted by quantum field theory combined with general
relativity.

\bigskip

\noindent We shall next amplify on the distinction,
and study more the entropy and surface screen considerations,
showing that even the present fraction of the
critical energy associated with acceleration can thereby
be understood.
The next portion derives an expression for the pressure, which is negative and thus a tension in the direction of the screen. 
The following results rely on simple principles of entropy and thermodynamics and are not dependent on any specific model.

\bigskip
\noindent The entropy on the Hubble Horizon, e.g. the Hubble radius $R_H = c/H$, is

\begin{equation} 
S_H = \frac{k_B c^3}{G \hbar} \frac{A}{4} = \frac{k_B c^3} {G \hbar} \pi R_H^2 = \frac{k_B c^3} {G \hbar} \pi \left( \frac{c}{H} \right)^2 \sim (2.6 \pm 0.3) \times 10^{122} k_B
\end{equation}


\noindent Increasing the radius $R_H$, by $\Delta r$, increases the entropy by $\Delta S_H$ according to

\begin{equation} 
\Delta S_H =  \frac{k_B c^3} {G \hbar} 2\pi R_H \Delta r = \frac{k_B c^3} {G \hbar} 2\pi \left( \frac{c}{H} \right) \Delta r \sim (2.6 \pm 0.3) \times 10^{122} k_B \Delta r / R_H
\end{equation}
The entropic force is simply
\begin{equation}
F_r = -  \frac{dE}{dr} = -  T  \frac{dS}{dr}  = - T_\beta \frac{ d S_H}{d r} = - \frac{\hbar}{  k_B }~\frac{ H }{ 2 \pi} \frac{k_B c^3} {G \hbar} 2\pi \left( \frac{c}{H} \right)  = - \frac{c^4 }{ G}  
\end{equation}
where the minus sign indicates pointing in the direction of increasing entropy or the screen, which in this case is the horizon.

\bigskip
\noindent The pressure from entropic force exerted is
\begin{equation}
P =  \frac{F_r}{A} = -  \frac{1}{A} T  \frac{dS}{dr}  = -  \frac{1}{A} \frac{c^4 }{ G}   =  - \frac{1}{ 4 \pi c^2/H^2}  \frac{c^4 }{ G}  = -  \frac{c^2 H^2 }{4 \pi  G} = -  \frac{2}{3} \rho_{critical} c^2 
\end{equation}
where $\rho_{critical}$ is the critical energy density $\rho_{critical}  \equiv 3 H^2 /8 \pi G$.

\bigskip
\noindent
This is close to the value of the currently measured dark energy/cosmological constant negative pressure (tension).
In this case the tension does not arrive from the negative pressure of dark energy but from the entropic tension due 
to the entropy content of the horizon surface.
This is equivalent to the outward acceleration $a_H = cH$ of Eq. (\ref{acceleration}).

\bigskip
\noindent If we chose to put the information screens at smaller radii, then, associating entropy with information,we would have found a proportionally smaller pressure, and an acceleration that decreases linearly with the radius in accordance with our expected Hubble law.
Thus, the acceleration of the universe simply arises as a natural consequence of the entropy on the horizon of the universe.

\bigskip

\section{Acceleration from the Entropy and Surface Terms}\label{sec:surt}

\bigskip

\noindent In this section, we present a specific phenomenological model inspired by surface terms usually ignored in general relativity  and show that this also leads to an accelerating universe. While the surface terms source our motivation, the model is best viewed as a phenomenological model which we
introduce here without rigorous derivation.
We consider the least additional assumption is that general relativity is correct, and that it can be easily understood and derived from a variational principle using the action. 

\bigskip
\noindent We show that, under reasonable assumptions, the new terms lead to an acceleration term in the Friedmann-Lema\^{i}tre equations. 
There is a solution to the acceleration equation that evolves from a decelerating to an accelerating phase. 
Our discussion of surface
terms presents a specific class of models that give rise to cosmological acceleration; however, we note that our main result, the derivation of cosmological acceleration as an entropic force presented in the previous section, is model independent. 

\bigskip
\noindent  The Einstein-Hilbert action including the surface term  and a matter action is (schematically)

\begin{equation}\label{actsur}
I =  \int_M \left(R +{ \mathcal L}_m \right) + \frac{1}{ 8 \pi}  \oint_{\partial M} K
\end{equation}

\bigskip

\noindent where $R$ is the scalar curvature, $\mathcal{L}_m$ is the matter and field Lagrangian, and $K$ is the trace of the extrinsic curvature of 
the boundary~\cite{HawkingHorowitz95}.
The application of variational procedures then produces the usual Einstein equations for general relativity with the addition of a surface energy term:
\begin{eqnarray}
{\rm Curvature ~of ~ Space-Time ~} & proportional & {\rm ~to ~the ~Stress-Energy ~Content ~ + ~Surface ~Terms }\nonumber \\ \nonumber \\
R_{\mu \nu} - \frac{1}{2} R g_{\mu \nu} & = & \frac{8 \pi G}{ c^4} T_{\mu \nu} + ~Surface ~ terms
\end{eqnarray}
Typically the surface terms are neglected though they make a significant appearance when a horizon is present.
This would in the case of spherical symmetry and homogeneity lead to the Friedmann-Lema\^{i}tre equations:
\begin{eqnarray}
{\rm Scale ~ factor ~acceleration}& = &{\rm  Energy ~Content ~deceleration ~ + acceleration ~from ~Surface ~Terms} \nonumber  \\ \nonumber \\
\frac{\ddot a}{ a} &=& - \frac{4 \pi G}{ 3} \left(\rho + \frac{3 P}{c^2} \right) +  a_{Surface}/d_H
\end{eqnarray}

\bigskip

\noindent Now, there are a number of approaches to determine the form of the new terms. Here we consider a simple possibility.
From our surface term motivation, we would anticipate that the integral of the trace of the intrinsic curvature would be of order $6 (2 H^2 + \dot H)$ so that the term would be approximately $\frac{6 (2 H^2 + \dot H)}{ 8 \pi} \sim \frac{3}{ 2 \pi}( H^2 +\dot H / 2)$.
We can also take an approach motivated by entropic ideas and see that these naturally lead to a slowly expanding, late time accelerating universe.

\bigskip

\noindent For a horizon, it is well known that there is an associated curvature and temperature, and that these two quantities are related. The temperature $T$ is given by the Unruh, de Sitter, or Hawking temperature prescriptions (except for a pesky factor of two difference between the Hawking temperature and the other two due to location of evaluation of the temperature - at the horizon or remote).
We can then associate the surface entropy or surface term with its temperature and its acceleration.
Using the relations we find (see Eq. (\ref{acceleration})) for the horizon acceleration
\begin{equation}
a_{surface} = a_{entropic} = c H
\end{equation}
where $H = \dot a / a$ is the Hubble expansion rate.
If we have a scale, which is naturally and necessarily the Hubble horizon scale, $d_H = c/H$, for our cosmological treatment, then we can complete the Friedmann-Lema\^{i}tre acceleration equation
\begin{equation}
\frac{\ddot a}{ a} = - \frac{4 \pi G}{ 3} \left(\rho + \frac{3 P}{c^2} \right) + H^2
\,.
\label{good}
\end{equation}
This is remarkably like the surface term order of magnitude estimate except for the $3/2 \pi$ factor.
With the Hawking temperature description the coefficient would have been 1/2. 
There is some freedom here and we chose the value that leads to nice equations in the two limiting cases.
It is easy to show that if the $H^2$ is highly dominant over the $  \frac{4 \pi G}{ 3} (\rho + 3 P/c^2)$,
the solution to the equation is simply a de Sitter space with scale factor  $a(t) = a(t_o) e^{H(t-t_0)}$.

\bigskip
\noindent The alternate equation
\begin{equation}
\frac{\ddot a}{ a} = - \frac{4 \pi G}{ 3} \left(\rho +  \frac{3 P}{c^2} \right) + \frac{3}{2 \pi} H^2  + \frac{3}{4 \pi} \dot H
\label{bad}
\end{equation}
may provide a better fit to the data (see \S{\ref{sec:dis}}) or a rigorous derivation but does not have the simplicity of Eq. (\ref{good}).

\bigskip
\noindent In order to adopt a broader approach, we generalized our discussion by considering
\begin{equation}
\frac{\ddot a}{ a} = - \frac{4 \pi G}{ 3} \left(\rho +  \frac{3 P}{c^2} \right) + C_H H^2  + C_{\dot H} \dot H
\label{general}
\end{equation}
where we anticipate the coefficients to be bounded by $ \frac{3}{2 \pi} \lsim C_H \leq  1$ and
$0 \leq C_{\dot H} \lsim \frac{3}{4 \pi} $.

\bigskip

\subsection{Comparison with supernova data}\label{sec:dis}
\begin{figure}
\centerline{\includegraphics[width=4.0in]{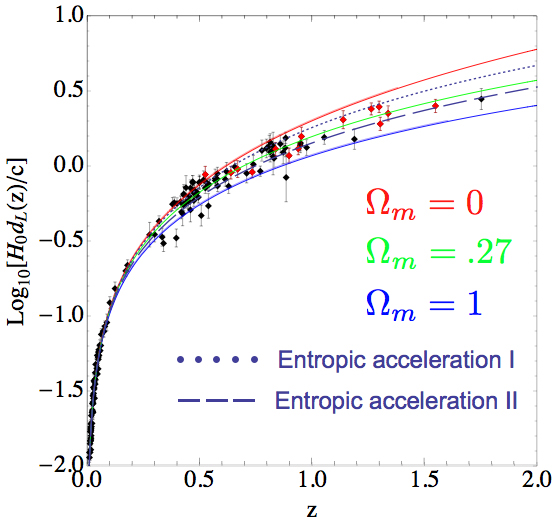}}
\caption{Comparison of entropic acceleration and several $\Lambda$CDM models.
The supernovae data points are plotted with error bars and the data is taken from~\cite{Riess:2004nr}.
The luminosity distance $d_L$ for the entropic models $I$ (Eq. (\ref{good})) and $II$ (Eq. (\ref{bad})) are denoted by the dotted and dashed (blue) curves respectively.
The theoretical predictions for $\Lambda$CDM  are represented by the solid curves.}
\label{dlfig}
\end{figure}

\noindent We conclude with a demonstration that the entropic acceleration mechanism inspired by surface terms can provide a surprisingly remarkable fit the supernova data, assuming the simple
form for the acceleration equation (\ref{bad}). Because
we are using a metric theory of gravity,
we may use the standard formula for the luminosity distance:
\be\label{lumd}
d_L(z;  H(z), H_0)  = \frac{c(1+z)}{H_0}  \int_0^z \frac{ dz'}{H(z')}
\,,
\ee
where $z$ is the redshift defined by $z + 1 \equiv a_0/a$.
For $\Lambda$CDM, the luminosity distance can be written~\cite{Perlmutter98}:
\be\label{lumlcdm}
d_L(z;  \Omega_M, \Omega_\Lambda, H_0)   = 
 \frac{c (1+z)}{H_0 \sqrt{| \kappa |}} \, {\mathcal S} \left( \sqrt{|\kappa|} \int_0^z dz' \left[ (1 + z')^2 (1 + \Omega_M z') -z'(2+z')\Omega_\Lambda \right]^{-\frac{1}{2}} \right)
\ee
where, $\mathcal S(x) \equiv \sin(x)$ and $\kappa = 1-\Omega_{tot}$ for $\Omega_{tot} >1$ while $\mathcal S(x) \equiv \sinh(x)$ with $\kappa = 1 - \Omega_{tot}$
for $\Omega_{tot} <1$ while $\mathcal S(x) \equiv x$ and $\kappa =1$, for $\Omega_{tot} =1$. For the  $\Lambda$CDM models we take $\Omega_{tot} = \Omega_M + \Omega_\Lambda$. Here we have defined $\Omega_M \equiv \rho_M/\rho_c = 8 \pi G\rho_M/3 H^2$ and  $\Omega_\Lambda =  \Lambda /3H^2$ where
$\rho_c$ is the critical energy density. The results are that the entropic acceleration models we consider can provide excellent fits to the
data as can be seen from Fig.~(\ref{dlfig}).~\footnote{In solving the equations for the fitting we have assumed
the standard scaling behavior for matter and radiation. This scaling behavior is now determined only in part by the Friedmann-Lema\^{i}tre equation and we do not
propose any novel form for this constraint yet; rather, we assume that the standard scaling applies
and should provide an adequate approximation for the purposes of Fig. (\ref{dlfig}).}   
The entropic models move smoothly from a decelerating to an accelerating phase
sometime near a redshift of $z=.5$, analogous to the $\Lambda$CDM models.
We suspect a complete model may be further constrained by consideration of Big-Bang Nucleosynthesis and possibly by precision
data relating to the equivalence principle. 

%

\section{Conclusions}

\noindent We have proposed a theory underlying the accelerated expansion
of the universe based on entropy and entropic force. This approach, while admittedly heuristic, 
provides a physical understanding of the acceleration phenomenon
which was lacking in the description as dark energy.
The evidence and general arguments supporting our hypothesis were presented in~\S \ref{sec:ent}. In addition we considered an interesting
phenomenological model, loosely based on surface terms in \S \ref{sec:surt}, and showed the models are capable of providing a good fit
to the supernova data. 

\bigskip
\noindent Following the above arguments to their logical conclusion, the accelerated expansion rate is the inevitable
consequence of the entropy associated with the holographic information storage on a surface screen placed at the horizon of
the universe. An interesting question~\cite{EFS2} is: how does this entropic viewpoint
of cosmic acceleration impact on inflationary theory?
\begin{center}

\section*{Acknowledgements}

\end{center}

\noindent We each thank our colleagues for their contagious energy and enthusiasm and IPMU for providing the venue that encouraged this work. This work was supported by the World Premier 
International Research Center Initiative (WPI initiative), MEXT, Japan. 
The work of D.A.E is supported in part by a Grant-in-Aid for Scientific Research
(21740167) from the Japan Society for Promotion of Science (JSPS), by funds from the Arizona State University Foundation
and by the National Science Foundation (KITP, UCSB) under Grant No. PHY05-51164.
The work of P.H.F. was supported in 
part by U.S. Department of Energy Grant No. DE-FG02-05ER41418. 
G.F.S. work supported in part by by the U.S. Department of Energy under Contract No. DE-AC02-05CH11231, by WCU program of NRF/MEST (R32-2009-000-10130-0),
and by CNRS Chaire Blaise Pascal.

\appendix*
\section{The Friedmann Equations from the 1st Law}
In this Appendix, we generalize our discussion to include the possibility that $\Omega \ne 1$.  We include a curvature term and use the apparent event horizon $r_A$  as our preferred screen rather than the Hubble horizon $r_H$--these are equivalent for the case of vanishing curvature. We have
\begin{equation} 
r_A = \frac{c}{\sqrt{H^2 + k / a^2}}   ~~~~~~{\rm and} ~~~~~~~~ \dot r_A = - \frac{H r_A^3}{c^2} \left(\dot H - \frac{k}{a^2} \right)
\end{equation}
The energy flow across the apparent horizon is $-dE = dM c^2 + P dV = (\rho c^2 + P) dV$
\begin{equation}
-dE = (\rho +P/c^2) c^2 dV = (\rho +P/c^2) c^2 A_A  v dt = (\rho +P/c^2) c^2 A_A  H r_A dt
\end{equation}
where $A_A = 4 \pi r_A^2$.
Assuming that the apparent horizon has an associate entropy $S$ and approximate temperature $T$ given by
\begin{equation}
S = \frac{k_B c^3}{\hbar G} \frac{A}{4}, ~~~~~~  T = \frac{\hbar H }{2 \pi k_B} = \frac{\hbar}{2 \pi k_B} \frac{c}{r_A} \,,
\end{equation}
we can use the first law of thermodynamics $-dE = T d S$ to find
\begin{equation}
-dE = A(\rho +P/c^2) c^2 H r_A dt = T dS = T \frac{k_B c^3}{4\hbar G} \frac{dA_A}{dt} dt =  T \frac{k_B c^3}{4 \hbar G} 2 \times 4 \pi r_A \dot r_A dt 
\end{equation}
Dividing by $c^2 dt$ and using $T = \frac{\hbar}{2 \pi k_B} \frac{c}{r_A}$ one finds and then substituting in for $\dot r_A$, 
\begin{equation}
A (\rho +P/c^2) H r_A = \frac{c^2}{G} \dot r_A =  - \frac{1}{G } H r_A^3 \left(\dot H - \frac{k}{a^2} \right)
\end{equation}
Dividing through by $H r_A / G$ and using $A = 4 \pi r_A^2$ we have  
\begin{equation}
4 \pi G (\rho +P/c^2) =  - \left(\dot H - \frac{k}{a^2} \right)
\end{equation}
Rearranging one has one form of the standard Friedmann acceleration equation
\begin{equation}
\dot H - \frac{k}{a^2} = - 4 \pi G (\rho +P/c^2) 
\label{Thermo1st}
\end{equation}

\noindent 
If one takes for the continuity (conservation) equation for the perfect cosmological fluids,
\begin{equation}
\dot \rho + 3 H ( \rho + P/c^2) = 0
\end{equation}
We can substitute $H(\rho + P/c^2)= -\dot \rho / 3$ into the equation for $\dot H$ and integrate
\begin{equation}
H \left( \dot H - \frac{k}{a^2} = - 4 \pi G (\rho +P/c^2) \right) =  \frac{4 \pi G}{3} \dot \rho
\end{equation}

\begin{equation}
\int H  \dot H - k \int \frac{\dot a}{a^3}  =  \frac{4 \pi G}{3} \int  \dot \rho
\end{equation}
Which yields
\begin{equation}
\frac{1}{2} \left( H^2 + \frac{k}{a^2} \right) = \frac{4 \pi G}{3} \rho + constant/2
\end{equation}
Or
\begin{equation}
 H^2 + \frac{kc^2}{a^2} = \frac{8 \pi G}{3} \rho + constant   \rightarrow H^2 = \frac{8 \pi G}{3} \rho -  \frac{kc^2}{a^2}+ \frac{\Lambda c^2}{3}
\end{equation}

This is simply the usual second Friedmann energy equation.
We can find the other form of the acceleration equation by using this equation and the relationship
$\dot H = \ddot a / a - H^2$ to substitute into Eq.~(\ref{Thermo1st}) to obtain
\begin{equation}
\frac{\ddot a}{a}  = - \frac{4 \pi G}{3}\left(  \rho + 3 \frac{P}{c^2} \right) + constant  \rightarrow \frac{\ddot a}{a}  = - \frac{4 \pi G}{3}\left(  \rho + 3 \frac{P}{c^2} \right) + \frac{\Lambda c^2}{3}
\end{equation}


\noindent
We can see from this derivation that the Friedmann equations naturally arise from the first Law of Thermodynamics and the association of Entropy and Temperature to the apparent horizon (or Hubble Horizon for a flat universe).
We also see there is a place for curvature (currently observed to be small) and for a constant of integration   $\Lambda$. 

\subsection{Continuity Equation from the First Law}
The continuity equation is derived assuming the simple form of the first law of Thermodynamics.
\begin{eqnarray}
d E & = & - P d V \cr
d M c^2 & = & - P d V \cr
d (V_0 \rho a^3) c^2 & = & - P d (V_0 a^3) \cr
d \rho c^2  a^3 + da  \rho c^2 3 a^2 &=& - 3 a^2 P da  \cr
d \rho &=& - 3 \left(\rho + \frac{P}{c^2} \right) \frac{1}{a} da \cr
\frac{d \rho}{d t} &=& - 3 \left( \rho + \frac{P}{c^2} \right)  \frac{1}{a}\frac{da}{dt} \cr
\dot \rho + 3 H  \left( \rho + \frac{P}{c^2} \right) &=& 0  
\end{eqnarray}

This suggests that our assumption of the standard continuity equation is appropriate for the comparison to data and in deriving the standard Friedmann equations.

\end{document}